# Boosting Light Absorption of a Therapeutic Microcapsule by Means of Auxiliary Solid Nanoparticles

## Yu.E. Geints[*], E.K. Panina

*V.E. Zuev Institute of Atmospheric Optics, 1 Acad. Zuev square, Tomsk, 634021, Russia*
*\*Corresponding author: ygeints@iao.ru*

**Abstract**

Multilayer microparticles with a liquid core and a polycomposite light-absorbing shell are important components of modern bio- and medical technologies, in particular, as transport microcontainers in the system of targeted delivery of therapeutic nanodoses to the desired region of the biological tissues. For reliably opening the microcapsule shell by an optical radiation and releasing the payload, it is necessary to dramatically increase the light absorption of such a microcontainer. To this end, we propose surrounding the microcapsule with specially added auxiliary nanoparticles, which can accumulate optical energy near its surface and direct a concentrated photonic flux to the target microcapsule thus leading to its booster heating. Using numerical finite-difference time-domain (FDTD) calculations, we simulate and examine the absorption dynamics of the near-infrared optical radiation in a spherical microcapsule surrounded by solid nanoparticles of different optical properties (metal, biocompatible dielectric). We show that due to light scattering on nanoparticles, the optical field superlocalization in the "hot regions" on the microcapsule surface occurs. The three-fold light absorption enhancement can be achieved due to the addition of nanoparticles.

## 1. Introduction

One of the important research field of modern chemical, biological and medical technologies is the development and manufacturing the miniature portable systems for encapsulation of various substances based on composite micron-sized particles – the microcapsules. There are several strategically important ways of microcapsule application [1-3] and, particularly, in medicine as a system of targeted delivery of biologically active components to cells and tissues. Ultimately, this allows for dramatically reducing the doses of delivered drugs and removing the side effects of their usage. Here, the principal challenge is not only the creation of microcapsules with specified properties but also the remote control on their transportation and controlled opening the capsule shell to release the payload, in particular, through optical radiation impact due to the photothermal effect [4, 5].

Currently, the control and management of the heat release inside an optically irradiated capsule is an outstanding scientific and practical task. On the one hand, there is a technical need to realize the highest possible efficiency of optical energy conversion into heat. On the other hand, too high heating of the transported substance in the microcontainer may be undesirable for its activation and subsequent functioning, which requires minimization of heat release in certain regions of the capsule. Additionally, the inhomogeneous distribution of the electromagnetic field and, hence, the temperature over the microcapsule surface can trigger its mechanical movement due to the photophoretic effect. To solve these problems, the detailed information on the spatial distribution of heat release regions and the temperature field inside the microcapsule irradiated with optical radiation is required.

Modern trends in the technology of microencapsulation of active substances are shifting toward the creation of multifunctional cargos that would be susceptible to various physical and chemical external stimuli for controlled release of the contents. Typically, to impart multifunctionality such capsules are manufactured in the form of micron-sized particles with a multilayered shell. The shell can be composed of several heterogeneous organic/inorganic layers responding to various external physic-chemical factors [6].

Generally, to make a microcapsule optically active in the visible and near-infrared spectral regions, the photothermal capsule activation is used by doping the shell with different substances strongly absorbing laser radiation, e.g., noble metal nanoparticles (gold, silver) [7], liquid laser dyes [8], or metal oxides [9]. Particularly, recently the technology of bifunctional microcapsule with polysaccharide shell decorated with inclusions of iron oxide nanoparticles ($Fe_3O_4$) and graphene is proposed which provides susceptibility of the microcapsule to both alternating magnetic field in the range of microwave frequencies (magneto-thermal effect) and optical radiation [10].

Importantly, the net absorbed optical power in a microcapsule depends not only on the absorption coefficient of shell material but on the optical field distribution in close proximity of the microcapsule also. Therefore, if one has managed to create the local areas of strong field concentration near the capsule shell (field focusing regions), then in these areas one could achieve a significant increase in the efficiency of light energy utilization into heat used for thermal activation of the microcapsule.

Possibly, one of the promising ways for the light absorption manipulation in microcapsules is using the auxiliary sol containing the nano-sized or micro-sized particles as a catalyst influencing the spatial distribution of heat release sources in capsule shell. Noteworthy, the use of nano-sized plasmonic particles in the form of coarse dope or disperse *nanofluid* is currently actively adopted in photovoltaics as an efficient scheme for capturing light radiation on solar cell surfaces, which contributes to better charge carrier transport [11-15]. Non-absorbing dielectric microparticles, e.g., silica ($SiO_2$), are also known to be used for this purpose producing directional light scattering and light concentration in thin-film cells [16] and operating as microlenses and Mie-cavities. It is also worth mentioning the example of such microlenses application in optogenetics to enhance the activation of specific light- sensitive membrane receptors (opsins) of neuronal and cellular bio-tissues [17].

In present work, for the first time to our knowledge, the optical absorption of the near-infrared optical radiation by a spherical microcapsule surrounded by specific auxiliary nanoparticles (NPs) with different optical properties is considered in detail. The auxiliary NPs serve to localize the optical energy near capsule surface and subsequent transfer it in the form of concentrated photonic flux which in the following leads to booster heating of the target microcapsule. A strongly absorbing metallic (gold) nanosol as well as non-absorbing nanospheres made of biocompatible material ($SiO_2$, $CaCO_3$, $TiO_2$) are considered as such a sol mediator. It turns out that the addition of a NP sol can increase net absorption by the microcapsule up to several times and this effect becomes stronger for larger NPs and for higher nanosol refractive index (if dielectric NPs are used).

## 2. Computer model of a microcapsule surrounded by nanoparticles

Numerical simulations of light scattering and absorption by a microparticle are carried out by solving a system of differential Maxwell equations for the electromagnetic field near the microcapsule using the finite-difference time-domain (FDTD) technique provided by Lumerical FDTD software. For adequately resolving the optical field near the metallic nanoparticles, the maximum spatial step of adaptive anisotropic numerical grid is set to 2 nm with a time marching of no more than 0.04 fs, which corresponds to approximately 100 mesh cells per wavelength. The boundaries of the computational domain are enclosed in a system of perfectly absorbing layers (PML) simulating the conditions of wave free propagation through the outer edges of the domain. The total number of Yee mesh cells could reach 50 million with an average running time of one variant of about 3 hours on an Intel$^R$ Core$^{TM}$ i7-7820X processor with 12 threads and memory allocation of about 12GB RAM.

A microcapsule is modeled as a two-layered spherical particle with a nonabsorbing water core simulating the payload and a light-absorbing solid-phase shell (Fig. 1a). Technical silicone doped with cylindric rod-shaped gold nanoparticles (nanorods) are used providing optical absorption in shell material. According to our previous calculation [18], the nanoparticles with such shape provide enhanced light absorption of a microcapsule shell in the near-infrared optical region (Fig. 1b). In the calculations, the capsule radius $R_c$ is 0.5 µm with a shell thickness 64 nm, which is about $0.08\lambda$ at optical wavelength $\lambda = 800$ nm. As shown in Ref. [19], this ratio of the shell thickness to the optical wavelength provides the maximum power density of light absorption inside a spherical microcapsule. The nanorods randomly distributed inside the shell give an absorption peak at λ=0.755 μm corresponding to the excitation of the longitudinal plasmon mode of such NP. Due to the chaotic spatial orientation of the nanorods with respect to the plane of light polarization, the spectral contour of the plasmon resonance broadens and microcapsule shell exhibits a smooth absorption maximum in the spectral region from 730 nm to 790 nm [18].

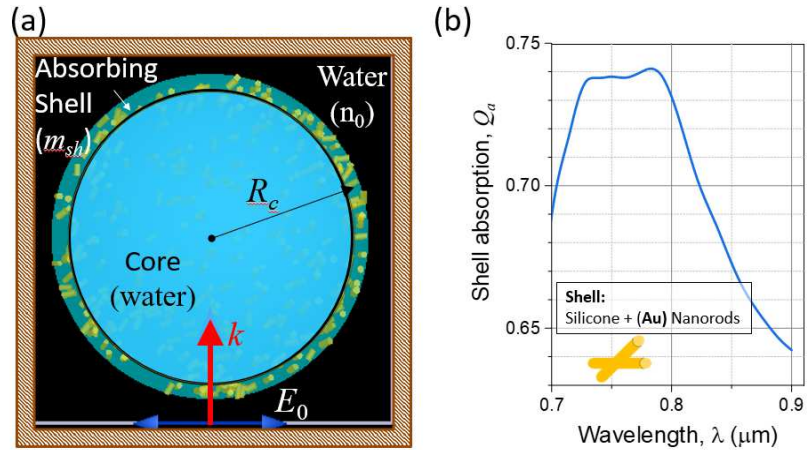

Fig. 1. (a) 2D-schematic of light-absorbing microcapsule as a two-layered sphere with a nanocomposite shell and water core; (b) Spectrum of optical absorption of a microcapsule with 10% gold nanorods volume fraction.

In the following, to simplify the calculations the microstructural composition of the capsule shell is not considered. Instead, a certain homogeneous medium with some effective complex refractive index $m_{sh}$ is adopted, which is calculated using the Brueggemann effective medium model with 10% volume fraction of nanorods in the shell. Thus, for selected laser wavelength of 800 nm (the main harmonic of the titanium-sapphire laser) one obtains the following value for effective refractive index of shell $m_{sh} = 3.6 - j\, 0.07$. The capsule core and the surrounding medium are considered as liquid water with refractive index $n_0 = 1.33$ and zero absorption of infrared optical radiation.

A plane linearly polarized optical wave with a unit electric field amplitude $E_0$ and a wave vector directed along the $x$-axis (shown by arrows in Fig. 1a) is set at the lower boundary of the region. Based on the calculations of the electromagnetic field distribution $\mathbf{E}(\boldsymbol{\rho})$ at each point with the radius-vector $\boldsymbol{\rho}$, the integral light absorption $P_a$ by a microcapsule is calculated as:

$$P_a = \frac{\pi c \varepsilon_0}{\lambda} \int_{V_c} d\mathbf{r}\, \varepsilon''(\boldsymbol{\rho}) |\mathbf{E}(\boldsymbol{\rho})|^2 = P_0 \int_{V_c} d\boldsymbol{\rho}\, q_a(\boldsymbol{\rho}) \qquad (1)$$

Here, $V_c$ is capsule volume, $\varepsilon_0$ is vacuum permittivity, $\varepsilon''(\boldsymbol{\rho})$ is the imaginary part of complex dielectric permittivity of capsule shell, $c$ is speed of light in vacuum, $q_a(\boldsymbol{\rho}) = \varepsilon''(\boldsymbol{\rho}) B(\boldsymbol{\rho})/2\lambda S$ stays for reduced absorption power density with $B(\mathbf{\rho}) = |\mathbf{E}(\mathbf{\rho})|^2 / E_0^2$ denoting field enhancement efficiency inside the capsule. Laser pulse power $P_0$ incident on the capsule midsection $S$ depends on the particle orientation relative to the wave polarization. As seen, the value of $P_a$ is mainly influenced by the heterogeneity of the optical field distribution in the microcapsule volume, which is accounted by the factor $B(\mathbf{\rho})$.

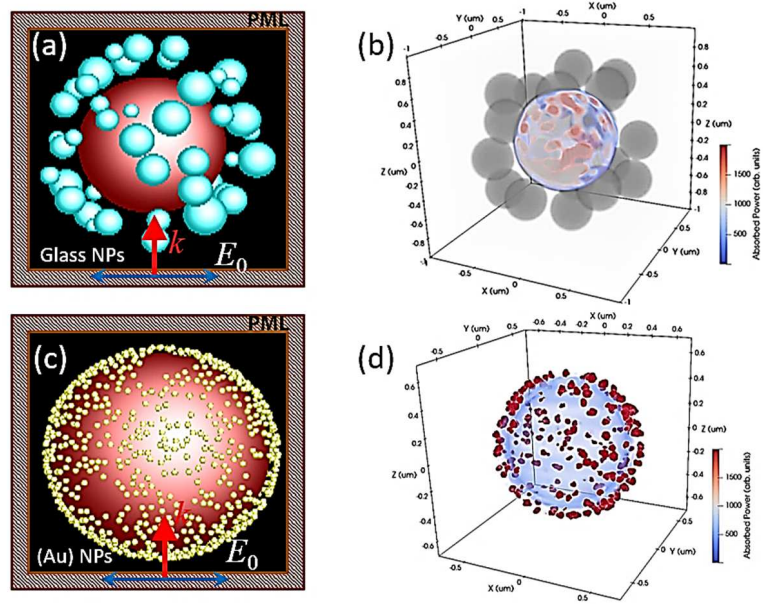

Fig. 2. (a, c) Model of a microcapsule with (a) dielectric and (c) gold auxiliary NPs. (b, d) 3D-distribution of absorbed power density $q_a$ on the surface of the microcapsule.

To deposit an ensemble of auxiliary (AUX) nanoscale scatterers around the microcapsule, a home-made software algorithm is used providing the generation of an array of NP centers distributed according to the random (normal) law. To this end, a globular layer encompassing the capsule is selected inside the computation domain with inner and outer radii $R_1 = (R_c + \gamma)$ and $R_2 = R_c + 2(r_{max} + \gamma)$, respectively, where $r_{max}$ is the maximum radius of the particle distribution function of NP ensemble, and $\gamma$ is a certain small parameter ($\gamma < r_{max}$). It is assumed that the NP array is entirely housed in this globular layer. Then, using the built-in random number generator, the Cartesian coordinates of $i$-th NP center, $\mathbf{\rho}_i = (x_i, y_i, z_i)$, is calculated in the loop, as well as the radius $r_i$ for each spherical NP. Two conditions are checked upon this procedure: (a) every new NP should be localized within predefined spherical layer, i.e., $R_1^2 \leq (x_i^2 + y_i^2 + z_i^2) \leq R_2^2$, and (b) every new NP should not overlap the neighboring nanoparticles which is controlled by calculating the distance between their centers: $\left[(x_i - x_m)^2 + (y_i - y_m)^2 + (z_i - z_m)^2\right] \geq (r_i + r_m)^2, \forall i, m$. If the center coordinates satisfy these conditions, the corresponding NP is rendered and becomes the set-up optical properties. As a final result, the microcapsule is surrounded by an ensemble of nanospheres randomly distributed directly near the capsule surface. Some examples of the capsule surrounded with the NPs in the form of dielectric and gold nanospheres are presented in Figs. 2(a, c). The calculated power distributions of absorbed optical radiation in the microsphere shell are shown in Figs. 2(b, d).

## 3.  Microcapsule absorption mediated by AUX NPs

### 3.1.  Optical absorption by a single NP

First, examine how an auxiliary nanoparticle modifies the near-optical field scattering of an incident electromagnetic wave. The numerical calculation results on the field intensity near a metallic and dielectric NPs of different sizes and refractive indices are shown in Figs. 3(a, b). In the insets to these figures, the 2D-distributions of the optical intensity are plotted when calculated for a 40 nm gold sphere with complex refractive index of $m = n+j\kappa = 0.154-j\,4.92$ and a 400 nm glass sphere with $n = 2$, respectively. NPs are assumed to be placed in water and illuminated by a monochromatic optical wave with a wavelength, $\lambda = 800$ nm.

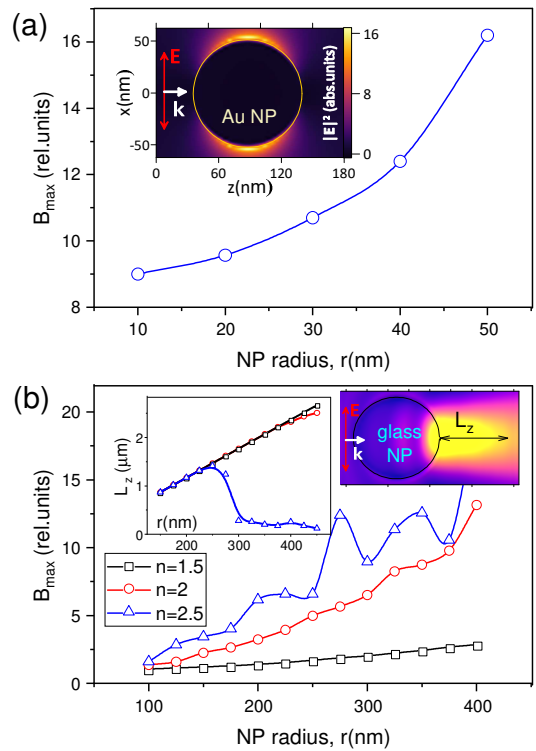

Fig. 3. Maximum optical field enhancement $B_{max}$ near (a) gold and (b) dielectric nanoparticles as a function of their radius ($r$) and optical properties ($n$) placed in water ($n = 1.33$). The insets show 2D profiles of optical intensity.

The analysis of the figures shows that in both cases nanoparticles concentrate the optical field near their surface. This effect is most pronounced for metallic (plasmonic) NP, where even at $r = 10$ nm two areas are formed with a peak intensity $B_{max} = 9$ and with a characteristic shape of dipole radiation in the direction of incident wave polarization. The length of these regions is of the order of the skin layer in metal and does not exceed 10 nm.

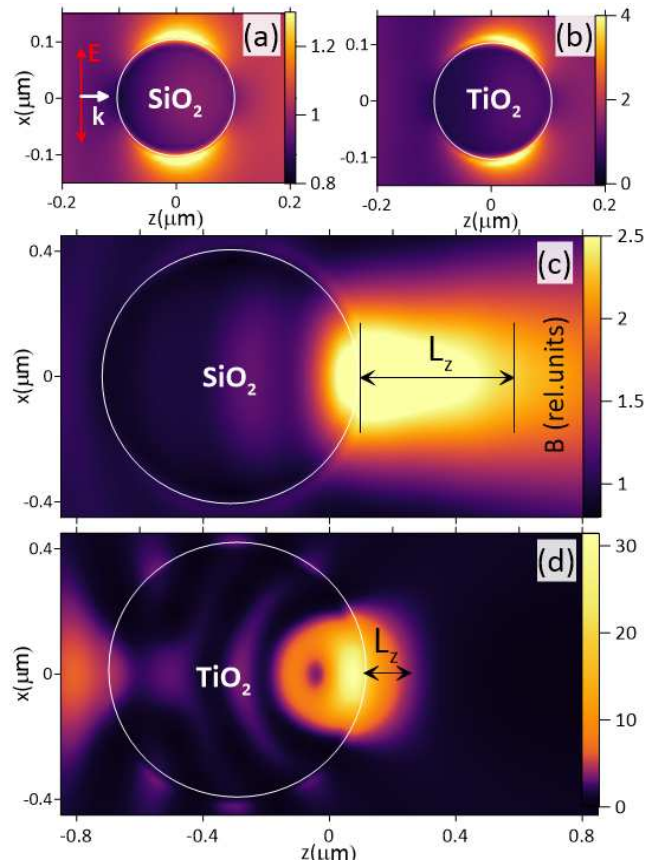

Fig. 4. (a-d) Color distributions of relative optical intensity near dielectric nanospheres with different size and refractive index: (a,b) $r = 100$ nm, (c,d) $r = 400$ nm; (a,c) $n = 1.5$ ($SiO_2$), (b,d) $n = 2.5$ ($TiO_2$).

In dielectric NPs, the character of the optical field distribution is completely different. Here, as seen in Fig. 4(c, d), a focal region with an increased intensity is formed near the shadow surface of the nanosphere. This region is often referred in the literature to as a "photonic nanojet" (PNJ) [20]. PNJ length $L_z$ is proportional to the particle size [21] and in the considered range of nanosphere radii is about one micrometer (see, Fig. 3b). It should be emphasized that the length of the PNJ is of particular importance here, because $L_z$ actually determines the spatial scale of the region where the optical intensity is still high due to NP field focusing.

Worth noting, in nanoparticles with a high refractive index, such as titanium dioxide ($TiO_2$) with $n = 2.5$ (Fig. 4d), the rear optical focus is shifted inside the particle, i.e., a PNJ starts from the particle inside [22]. This effect significantly shortens the outer PNJ part toward sub-nanometer values but the field intensity enhancement in this case remains high enough, $B_{max} \approx 20$. In Rayleigh-size dielectric nanospheres with $r \ll \lambda$ (Fig. 4a,b), an intensity distribution similar to that of metallic NP is formed, i.e., the field maximums are localized at the surface along the polarization vector of the incident wave. However, the focusing properties of such particles are poor, $B_{max} \approx 4$.

### 3.2. Microcapsule surrounded by NPs

### 3.2.1. Dielectric AUX NPs

For simplicity, we first consider the case when only one non-absorbing dielectric nanoparticle is placed next to the absorbing microcapsule. This situation is illustrated in Fig. 5 as 2D-profiles of the normalized optical intensity $B$ and the absorbed optical power $q_a$ in the microcapsule shell. Similar distributions for a single microcapsule (without a NP) are also shown here for comparison.

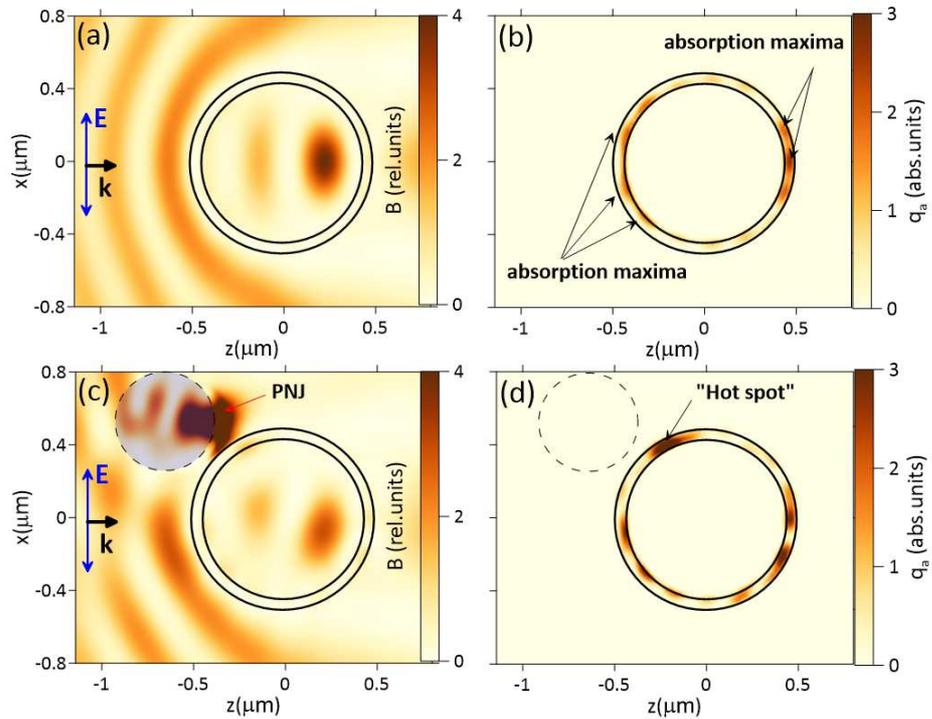

Fig. 5. 2D-distributions of (a, c) relative intensity $B$ and (b, d) absorbed power $q_a$ near (a, b) a single microcapsule and (c, d) particle dimer composed by a 300 nm $TiO_2$ NP with $n = 2.5$.

As seen, in the absence of NPs (Fig. 5a,b) the light absorption by the microcapsule is mainly realized in local areas on the illuminated and shadowed surfaces. If an additional NP is placed close to the capsule (Fig. 5c), a region with increased intensity resembling a PNJ appears near its shadow hemisphere. This region in its turn forms an additional "hot spot," which in this case is seen on the upper side surface of the microcapsule (Fig. 5d) where the absorption maximum is realized. Meanwhile, a NP imposes the distortion of the field scattering pattern, so the spatial distribution of absorption regions throughout the microcapsule shell also changes. Obviously, if there are many NPs, the absorption structure will be rather complex and it will be generally determined both by the number and mutual arrangement of NPs as well as on their optical activity (refractive index).

In order to determine these complex absorption dynamics, we perform a series of large-scale numerical calculations upon varying NP number $N$, NP median radius $r$, and refractive index $n$. In the following, instead of NP number it is more convenient to use the NP volume fraction parameter $\delta V$ in the globular layer surrounding the capsule, where the nanoparticle center is generated: $\delta V = \left(R_2^3 - R_1^3\right)^{-1} \int_{r_{min}}^{r_{max}} f(r) r^3 dr$, where $f$ is the size distribution function of the nanoparticles. The maximal intensity enhancement factor $B_{max}$ realized on the capsule surface, the peak absorption density $q_{max}$ and the total power of optical radiation $P_a$ absorbed by the microcapsule (which is calculated by the Eq. (1)) are chosen as key parameters of the microcapsule absorption activity. For clarity, the parameters $q_{max}$ and $P_a$ are normalized to the corresponding values realized in the capsule without AUX NPs: $\bar{q}_{max} = q_{max}/q_{a0}$ and $\bar{P}_a = P_a/P_{a0}$, where $q_{a0} = 2.55$ fW/μm$^3$, $P_{a0} = 0.134$ fW. The above-mentioned key parameters are shown in Figs. 6(a-d). In the calculations, NP size distribution is assumed monodisperse. Each point in these plots is obtained by statistical averaging over 50 independent program runs, giving a random NP distribution near the microcapsule. The average scatter of the obtained values is maximal for the parameters $B_{max}$ and $q_{max}$ and can reach 20% for small fractions of the bulk NPs content.

Recall that in the numerical calculations the amplitude $E_0$ of incident optical wave is set as 1 V/m, which gives initial optical field intensity of about ~ 0.17 μW/cm$^2$ and the laser power incident on the capsule median cross section of $P_0 = 1.33$ fW. In real situation, the optical power may be significantly higher and reach tens of kW/cm$^2$ [23]. This, in turn, will lead to a significant increase in the power of thermal sources formed in the microcapsule shell during absorption of optical radiation.

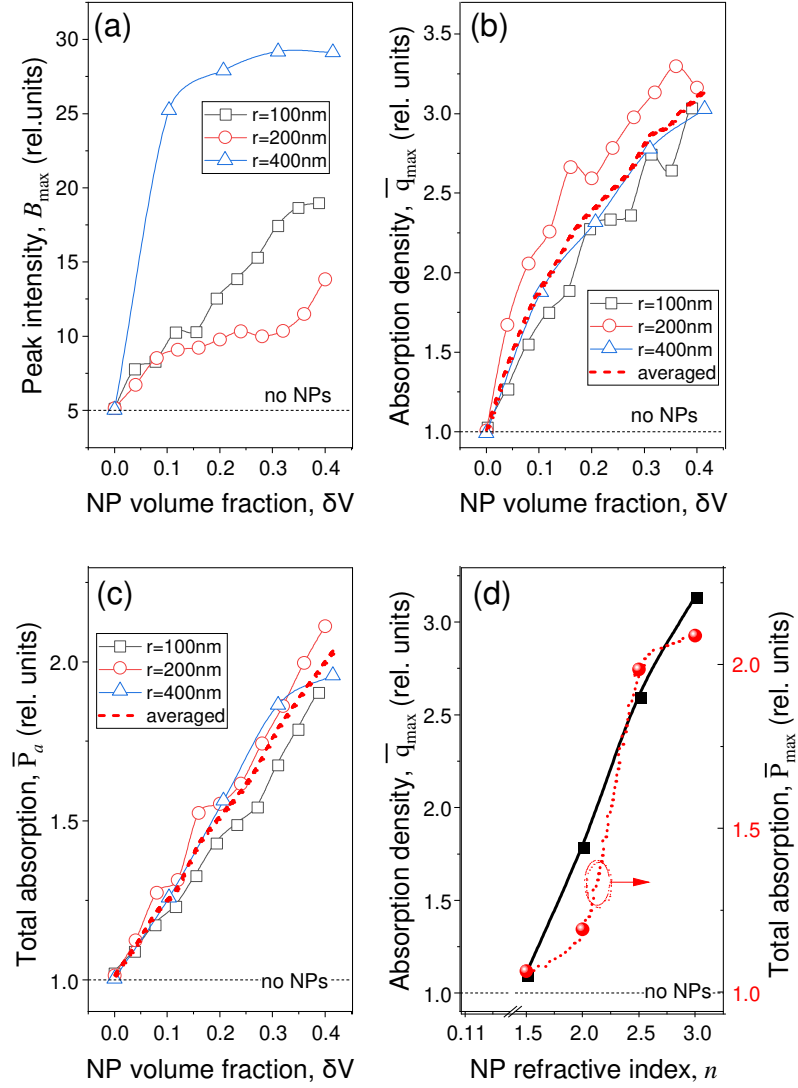

Fig. 6. (a-c) Optical absorption parameters of a microcapsule surrounded by TiO$_2$ NPs at the variation of NP volume fraction δV. (a) Maximum intensity enhancement $B_{max}$, (b) normalized peak absorption power density $\bar{q}_{max}$, (c) normalized total absorbed optical power $\bar{P}_a$. (d) Peak parameters of capsule absorption mediated by dielectric NPs with maximal possible volume fraction, δV = 0.34, and different refraction index $n$.

Fig. 6(a) shows the maximum possible enhancement of the optical intensity $B_{max}$ at the microcapsule shell, which can be achieved due to titanium dioxide NPs ($n$ = 2.5) addition near the mother particle. Clearly, that the growth in the NP number (volume fraction δV increase) and in NP size ($r$) causes corresponding increase of the optical intensity focused by NPs directly on the microcapsule. At the same time, bigger NPs lead to earlier saturation of $B_{max}$(δV) dependence. The physical cause is that relatively big nanoparticles achieve dense packing at smaller volume fraction, and further increase in their number leads to partial shielding of the microcapsule from the incident optical radiation by the outermost NPs, which are not in close contact with the capsule surface.

An increase in the optical intensity $B_{max}$ causes an increase in the local absorbed power in the "hot spots" of the microcapsule shell and a corresponding gradual grows of the total absorption power that can be clearly seen in Figs. 6(b) and (c). Interestingly, the scatter in the values of the parameters $q_{max}$ and $P_a$ with changes in the size of the auxiliary NPs turns out to be relatively small, which allows one fitting these curves by a single linear dependence shown in Figs. 6(b) and (c) with the red dashed lines.

From Fig. 6(d) it follows that the maximum absorption enhancement $\bar{q}_{max}$ in the "hot spots" achieved with the addition of transparent AUX NPs (relative to the value without NPs) increases with the increase in the optical density of nanoparticles. For silica ($SiO_2$) NP aerosol with $n = 1.5$, the absorption density practically does not change because the NPs only slightly focus the optical field (Fig. 3b). Contrarily, the absorption increases almost up to three times if titanium dioxide NPs with $n = 2.5$ are added near the microcapsule. Noteworthy, the maximum total power $\bar{P}_{max}$ absorbed by a capsule demonstrates more than double increase above its value without NPs.

### 3.2.2. Metal NP ensemble

Now recall Fig. 3(a), where the degree of optical field enhancement in the vicinity of gold nanoparticles is shown. As noted above, such metal nanoparticles can increase the optical intensity many times in two diametrically opposite spatial regions located near their surface. However, the length of these regions usually does not exceed tens of nanometers which requires close contact with NP surface to achieve effective interaction with mother capsule. At the same time, the small size of metal NPs makes it possible to increase the number of AUX NPs deposited near the microcapsule that can, on the contrary, enhance light absorption.

The dependence of the absorbed optical power in the microcapsule shell on the volume fraction of gold NPs of different radii (monodisperse sol) is shown in Fig. 7. Again, the power absorbed is normalized to its value for a single capsule, i.e., at $\delta V = 0$.

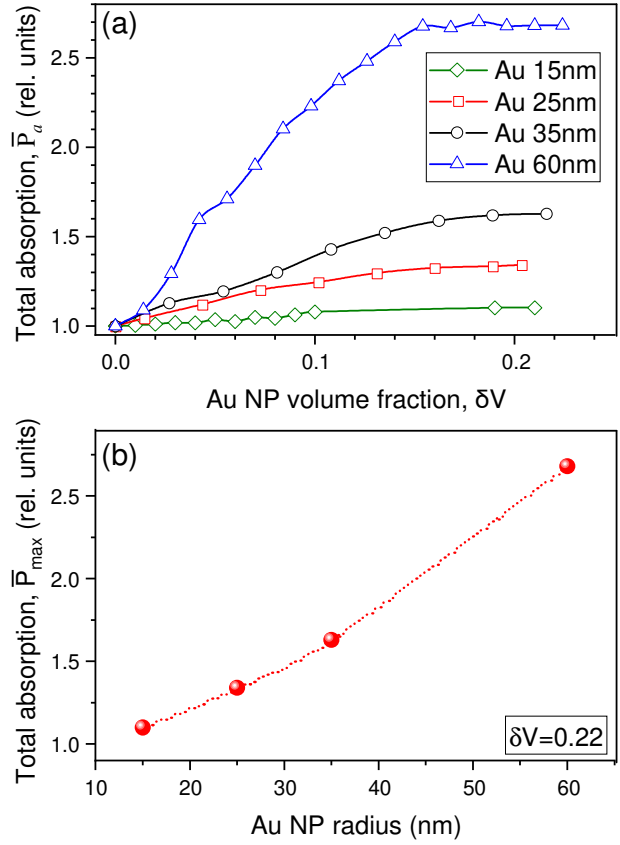

Fig. 7. (a) Optical power $\bar{P}_a$ absorbed by the microcapsule surrounded by gold AUX NPs of different radii and volume fraction $\delta V$. (b) Peak values of absorbed optical power $\bar{P}_{max}$ in the microcapsule bounded by Au NPs with different sizes at $\delta V = 0.22$.

As seen, according to the calculations presented in Fig. 3a the auxiliary nanosol containing larger metal NPs provides better light absorption by the microcapsule due to greater amplification of the optical field near NP rim. Thus, 10-nm Au particles practically do not change the absorption characteristics of the capsule, 15-nm NPs already give about 30% increase in $P_a$, and an even larger sol with $r = 60$ nm can increase the absorbed power almost threefold at the maximum concentration of AUX NPs (Fig. 7b).

Interestingly, the very tendency of $P_a$ increasing with increasing the NP volume content $\delta V$ alters with growing the NP size. The smallest gold nanoparticles considered lead to a monotonic and almost linear increase in the microcapsule absorption with the increase of their volumetric content in the surroundings. Meanwhile, the use of the largest NPs (60 nm) causes a rapid increase in total capsule absorption at low and moderate $\delta V$ values, but starting from $\delta V \approx 0.16$ this rise stops reaching the saturation level. Eventually, this may be due to the manifestation of partial shielding of the microcapsule surface by large metallic NPs, which unlike small nanoparticles effectively scatter the incident optical radiation and create multiple shadows on the capsule shell. Since not all NPs are closely adjoined to the microcapsule surface (because of their random generation), the effect in amplifying the near-surface field upon the volume fraction growth of NPs is less than one could expect.

Noteworthy, in our simulations the excitation of plasmonic resonances of the optical field in metal NPs is not addressed since the spectral range of surface plasmon manifestation in gold nanospheres lays in the visible optical band, approx. from 500 nm to 600 nm. At the same time, by changing the NP shape, e.g., using nanorods [18, 24] or nanostars [25] instead of nanospheres, one can shift the resonant frequency of the surface plasmon to the near-infrared range and realize the resonant amplification of the optical field. However, this giant field enhancement can cause a side effect because manifests itself in the inevitable heating of the nanoparticle and adjacent regions of medium. In turn, this may be prohibitive for therapeutic purposes. With these considerations in mind, the use of non-absorbing dielectric AUX NPs to boosting and control the optical absorption of the microcapsule looks safer and more promising.

## 4. Conclusion

To summarize, we theoretically address the problem of optical radiation scattering on a micron-sized bilayer spherical particle with a water core and an absorbing polycomposite shell. Such a photonic structure simulates a spherical microcontainer, a microcapsule, widely used in biomedical and pharmaceutical studies for transportation and targeted delivery the active substance to address areas. Because the microcapsule is assumed to be opened by an optical radiation, the aim of our research is increasing the light absorption of the microcapsule shell by the field concentration near particle surface. To this end, we propose using an auxiliary sol of dielectric or metallic (gold) nanoparticles distributed randomly near the microcapsule. The physical cause of the growth in microcapsule absorption stems from both the optical field focusing by a transparent nanoparticle at the nanoscale leading to the formation of a photonic nanostructure (a photonic nanojet), and the collective electron oscillations on the surface of plasmonic nanoparticles.

Using the FDTD numerical calculations, we demonstrate that in both cases (dielectric or metal nanosol) a considerable increase in the microcapsule absorption is realized. Importantly, the larger are the auxiliary sol nanoparticles and the higher is their refractive index (in the case of dielectric NPs), the stronger is the NPs influence on the capsule absorption parameters. In this case, the effect from the presence of gold NPs can be even stronger than that of a dielectric sol and under certain conditions can reach several times.

**Funding.** The authors acknowledge the financial support of the Russian Science Foundation (23-21-00018).